\documentclass[10pt,letterpaper]{article}
\usepackage{acl2015,indentfirst,times,secdot}
\usepackage{xspace,empheq,fancybox,amsmath,bbm,amssymb,epsfig,subfig,syntonly,times,amsthm,graphicx} \usepackage{psfrag,color,bm,array,cite}
\usepackage{url,cite,footnote,xspace,syntonly,algorithm,algpseudocode}

\usepackage{verbatim,multirow,diagbox,enumitem,tabu,acl2015}
\usepackage[T1]{fontenc}
\usepackage{setspace}
\usepackage[justification=justified,font={small,sf}]{caption}

\usepackage[colorinlistoftodos]{todonotes}

\input{mysymbol.sty}

\newcommand{\T}{\mathsf{T}}

\begin{document}
	
\title{\vspace{7.5mm} Enhancing the Spatio-Temporal Observability of Residential Loads}
\author{\normalsize \normalfont
 Shanny Lin and 	Hao Zhu \\
	\normalsize  Department of Electrical and Computer Engineering \\
	\normalsize The University of Texas at Austin \\
	\normalsize  Emails: \{shannylin,haozhu\}@utexas.edu}

 \maketitle

\begin{abstract}
Enhancing the spatio-temporal observability of residential loads is crucial for achieving secure and efficient operations in distribution systems with increasing penetration of distributed energy resources (DERs). This paper presents a joint inference framework for residential loads by leveraging the real-time measurements from distribution-level sensors. Specifically, smart meter data is available for almost every load with unfortunately low temporal resolution, while distribution synchrophasor data is at very fast rates yet available at limited locations. By combining these two types of data with respective strengths, the problem is cast as a matrix recovery one with much less number of observations than unknowns. To improve the recovery performance, we introduce two regularization terms to promote a low-rank plus sparse structure of the load matrix via a difference transformation. Accordingly, the recovery problem can be formulated as a convex optimization one which is efficiently solvable. Numerical tests using real residential load data demonstrate the effectiveness of our proposed approaches in identifying appliance activities and recovering the PV output profiles.
\end{abstract}
\vspace*{6pt}

\section{Introduction}
\label{sec:intro}

Power distribution systems are known to lack in real-time observability, especially for the individual residential loads. Limited amount of sensor measurements are available for distribution system monitoring, typically from a few line monitors and control devices. Consequently, traditional distribution state estimation (DSE) methods \cite{Primadianto2017review,Lu1995dsse,Baran1994se} suffer from low estimation accuracy and robustness, and, furthermore, they fail to provide timely visibility of residential loads.


To address this challenge, advanced sensing and communication technologies have been increasingly deployed in distribution systems. One type of sensing device is the distribution synchrophasor measurement unit (D-PMU) \cite{Meier2014}. D-PMUs provide high-quality phasor and power measurements at sub-second sampling rates of a few grid locations. In addition, smart meters are widely installed at every household to collect the electricity consumption data at intervals of 15 minutes to one hour \cite{AMI}. These two types of sensors show the trade-off between spatial diversity and temporal resolution. Ubiquitously available smart meter data can lose the transient load information due to time averaging, while the high-rate D-PMU data suffers from limited deployment due to high installation costs. Therefore, neither type of sensors alone can directly provide the dynamic load profile at each feeder node. 


Meanwhile, increasing penetration of distributed energy resources (DERs) greatly challenges distribution system management, thereby calling for enhanced spatio-temporal observability of residential loads. For example, to validate the security of electric vehicle (EV) charging command, one can estimate the start/end time of EV charging using the change points of the high-rate household load profile \cite{Niddodi2019ev}. Similarly, the dynamic load profile can also be used to verify residential photovoltaic (PV) systems' inverter control  settings \cite{Jacobs2019pv}. In addition to that, the real-time information on residential PV outputs is necessary for achieving effective protection design in distribution systems, as the solar generation level can significantly affect the fault current magnitude therein \cite{Zheng2018relay}. Hence, residential load profiles of high spatial and temporal resolutions are crucial for achieving secure and reliable operations in distribution systems. 


This paper proposes a joint inference framework for residential loads by leveraging the respective strengths of  D-PMU and smart meter data. To recover the spatio-temporal load matrix, both the smart meter data and the aggregated load measurements provided by D-PMUs are modeled as linear transformations of the load matrix (Sec. \ref{sec:sys_mod}). 
To tackle the identifiability issue 
therein, we first present two key characteristics of the underlying load matrix. While spatial correlation among load profiles manifests in a \textit{low-rank} matrix of lower-dimensional representation, the nodal appliance activities lead to \textit{sparse changes} in the matrix rows. Accordingly, the problem boils down to recovering a low-rank plus sparse matrix, closely related to the robust principal component analysis (RPCA) work; see e.g., {\cite{candes2011robust,Xu2012outlier,wright2009robust}}. Note that similar approaches have been popularly used in other applications in power systems such as synchrophasor data recovery \cite{Gao2016}, load data cleansing \cite{Mateos2013}, and forced oscillation location \cite{XieLe}. However, our residential load data matrix slightly differs from these existing models as a difference transformation is needed for achieving the low-rank plus sparse structure. Similar to earlier approaches, we introduce two convex regularization norms to match such structural characteristics and to cast the recovery problem as a convex optimization one (Sec. \ref{sec:recovery}). A post-processing procedure is developed to improve the estimation error bias due to the regularization penalty, while the recovery performance is discussed in relation to RPCA results (Sec. \ref{sec:perf}). Numerical tests using real residential load data demonstrate the effectiveness of our approaches in identifying large appliance activities and recovering the PV outputs (Sec. \ref{sec:simulation}). The presence of periodic HVAC loads has resulted in some performance degradation, thereby pointing out a future direction of including more diverse types of measurements (voltage/current/reactive power) for improved recovery performance.   


\textit{Notation:} Upper (lower) boldface symbols stand for matrices (vectors); $(\cdot)^{\T}$ stands for matrix transposition;  $\|\cdot\|_*$ denotes the matrix nuclear norm; $\|\cdot\|_1$ the L1-norm; $|\cdot|$ the entry-wise absolute value; and  $\mathbf 1$ stands for the all-one vector of appropriate size.


\section{System Model}
\label{sec:sys_mod}

\begin{figure}[tb] 
\centering
\vspace{-3pt}
\includegraphics[width=\linewidth,clip = true]{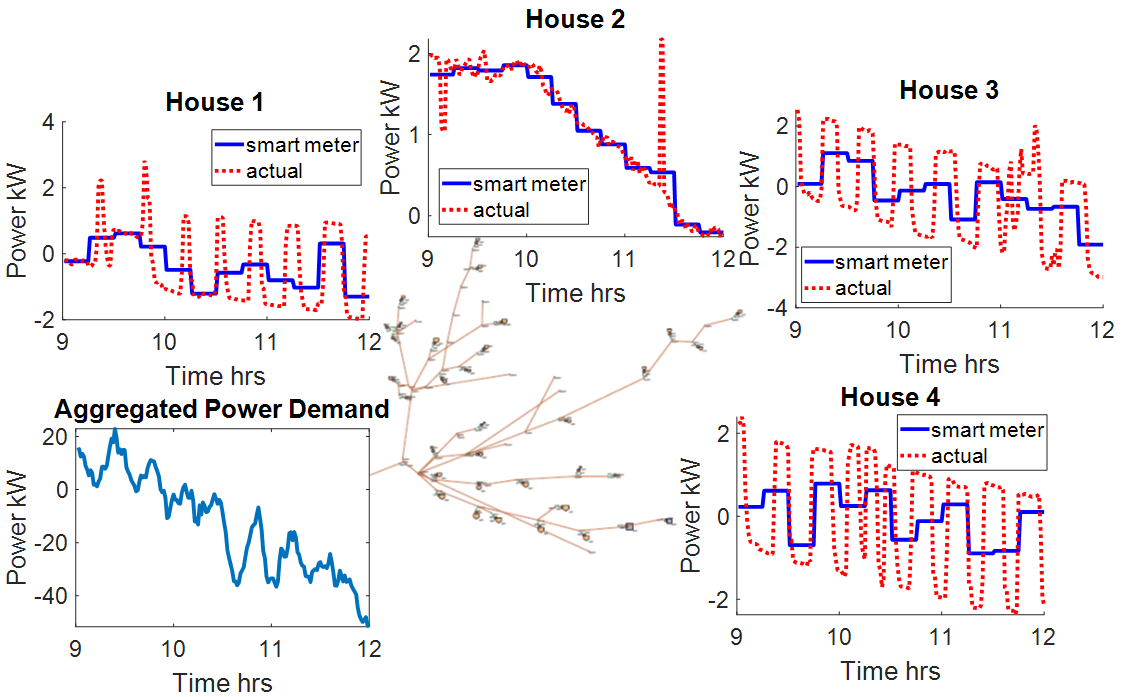}
\vspace{-4mm}
\caption{Overview of the distribution feeder system with multiple load nodes and various types of measurements.}
\label{fig:sys_overview}
\end{figure}

Consider a distribution feeder system with residential homes connected at the feeder ends as shown in Fig. \ref{fig:sys_overview}. The goal is to recover the spatio-temporal active power load matrix $\bbP \in\mathbb{R}^{N\times T}$ where $N$ is the number of load nodes (residential households) and $T$ is the total number of time slots. The temporal resolution of $\bbP$ represents the fastest time-scale of the all measurements. This work assumes a minute-level time resolution, which can be generalized to even faster time-scales such as the second-level time resolution of actual D-PMUs \cite{PSL}. 

To recover $\bbP$, we consider two types of measurement data, namely the smart meter data for each load node and the aggregated power demand at 
the feeder head. 
Typically, all residential households are equipped with smart meters that collect electricity consumption data at intervals of 15-minutes or one hour \cite{AMI}. Note that in this work, smart meter data is assumed to be available every 15 minutes by averaging the corresponding 15 samples in $\bbP$. For each house, every smart meter measurement recorded is the average active power consumed over the data collection interval. Given the 1/15 down-sampling rate, 
the smart meter data matrix $\bbY \in \mathbb{R}^{N\times T_s}$ with $T_s=T/15$ is given by  
\begin{align}
\bbY = \bbP\bbA+\bbE_Y  \label{sm_eqn},
\end{align}
where matrix $\bbA \in \mathbb{R}^{T\times T_s}$ represents the time averaging operation on $\bbP$ while $\bbE_Y$ denotes the measurement noise matrix.

At the aggregation location, a D-PMU can collect the total load profile, in addition to voltage/current phasors, with fast minute-level resolution and 
high quality. 
To simplify the model, we assume that the network losses are omitted from this aggregated measurement. Hence, the total load measurement $\bbz \in\mathbb{R}^{T}$ by aggregating over all $N$ houses is given by 
\begin{align}
\bbz^{\mathsf{T}} = \mathbf{1}^{\mathsf{T}}\bbP+\bbe_z^\T \label{pmu_eqn},
\end{align}
where $\bbe_z$ is the D-PMU measurement noise vector and the all-one vector $\mathbf{1}$ sums up all nodal profiles. Note that 
multiple D-PMUs can be included as well using a general matrix to replace $\mathbf{1}^\mathsf{T}$. 
Moreover, although we consider a simple aggregation scheme, the model in \eqref{pmu_eqn} can be generalized to include feeder losses as well. If the losses are a fixed percentage of the total demand, then one can scale the aggregated load measurement to reflect the consumed power only. The most general solution will be to represent the exact losses using (possibly linearized) distribution power flow models; see e.g., \cite{bernstein2018load}. 

Clearly, using the measurements in \eqref{sm_eqn} and \eqref{pmu_eqn}, the problem of recovering matrix $\bbP$ is underdetermined. The total number of equations, given by \eqref{sm_eqn} and \eqref{pmu_eqn}, equals to $(NT_s +T)$ which is much smaller than $NT$, the number of unknowns. Therefore, we will exploit certain characteristics and special structures of matrix $\bbP$ to achieve good recovery results. 

\section{Spatio-Temporal Load Recovery}
\label{sec:recovery}

\begin{figure}[tb] 
\centering
\vspace{-3pt}
\includegraphics[width=\linewidth,clip = true]{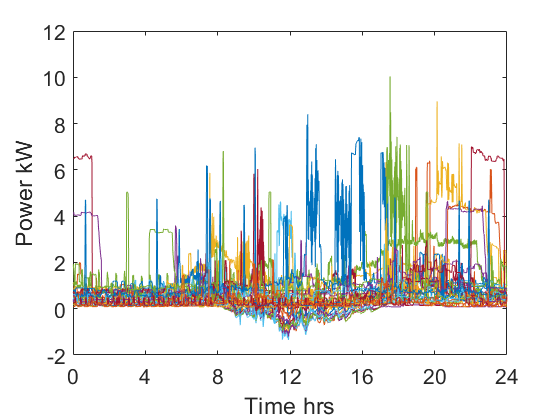}
\centerline{(a)}
\includegraphics[width=\linewidth,clip = true]{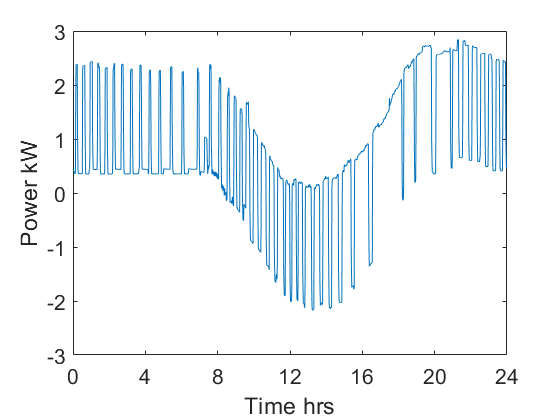}
\centerline{(b)}
\vspace{-4mm}
\caption{Actual residential load profiles available from the PecanStreet Dataport \cite{pecan} for (a) 30 houses on a winter day and (b) a single house with PVs on a summer day.}
\label{fig:P_observed}
\end{figure}

It is well known that load demand curves at the transmission level exhibit high correlation among nearby locations, and thus share similar \textit{temporal patterns}. This property has been widely used by load forecasting and data cleansing works; see e.g., \cite{wang2016electric,Mateos2013,Kim2013,Fiot2018}. Several factors play a role in leading to this similarity, including weather conditions (i.e. temperature, irradiance) and economic conditions (i.e. electricity prices). Although this spatial correlation property is well known to hold for transmission-level loads, we have observed it for distribution-level loads as well. 
Fig. \ref{fig:P_observed}(a) plots the load profiles of one winter day for 30 residential houses (15 of which have PV panels) located in the same neighborhood in Austin, TX. Fig. \ref{fig:P_observed}(b) plots the load profile of one selected house with PV during a summer day. Compared to the winter profiles, the summer one has a high level of periodic HVAC loads. These plots have been generated from minute-level real data available through the PecanStreet Dataport \cite{pecan}. It has been observed that there exists a daytime temporal pattern among 
the 15 houses with PVs, corresponding to a typical daily solar irradiance profile in Austin, TX. Similarly for the 15 houses without PVs, they share the same minimal base loading pattern.
Additionally, during the time periods with no solar irradiance, all 30 houses share a similar minimal base-load pattern. Thus, the spatial correlation among minute-level residential loads is mainly due to the PV output and the base loading, not from the usage of electric appliances.
We assume load matrix $\bbP$ has an underlying low-rank component $\bbL\in\mathbb{R}^{N\times T}$, rows of which are either highly correlated (PVs) or close to zero (no PVs).
Note that the load nodes  are located within the same feeder (connected to the same D-PMU), and, therefore, in the same neighborhood. This ensures that houses with PVs will exhibit similar solar irradiance patterns. The effect of a variable type of houses, including houses not co-located, on the similarity of temporal pattern and recovery performance will be investigated in future. 

Interestingly, residential load curves go beyond the temporal similarity of transmission loads as they also include rectangular waveforms which are not synchronized across locations. These components reflect the large appliance activities at individual households. For example, the visible ones in Fig. \ref{fig:P_observed}(a) correspond to the charging events of household electric vehicles (EVs), while the frequency patterns in Fig. \ref{fig:P_observed}(b) relate to the summer-time HVAC loads. Generally speaking, these appliance activities still occur infrequently over the course of a day and show no strong correlation with other households. Hence, they can be captured by \textit{sparse changes} in the daily load profiles, represented by an additional component $\bbS\in\mathbb{R}^{N\times T}$ in matrix $\bbP$. This sparse-change characteristic has been exploited by \cite{bhela2018enhancing} to tackle the identifiability issue when only a subset of loads are metered.
Note that $\bbS$ itself may not be sparse but rather piece-wise constant. Hence, the consecutive differences for each load are sparse, as defined by $D_{n,t} = S_{n,t}-S_{n,t-1}, ~\forall (n,t)$ with $S_{n,t} = \sum_{\tau=1}^t D_{n,\tau}$. Thus, matrix $\bbS$ becomes sparse under the linear transformation given by $\bbS=\bbD\bbU$  where $\bbU\in\mathbb{R}^{T\times T}$ is an upper triangular matrix of all ones.  

This way, the data matrix $\bbP$ can be decomposed into a low-rank matrix augmented by an additional sparse-change matrix such that $\bbP=\bbL+\bbS=\bbL+\bbD\bbU$. 
This formulation of $\bbP$ mimics the model used by the framework of robust principal component analysis (RPCA) which  decomposes a large data matrix into a low-rank 
plus sparse 
from \cite{candes2011robust}. For RPCA, the low-rank 
component similarly captures the spatial correlation among the rows, while the sparse component
represents outlier entries that are not correlated across rows. Hence, our data model is slightly different from the RPCA setting. However, by defining matrix $\bbW = \bbU^{-1}$ that can compute the consecutive differences between the columns, 
we obtain 
\begin{align}\label{matX}
\bbX:=\bbP\bbW=(\bbL+\bbD\bbU)\bbW = \bbK+\bbD
\end{align} 
where $\bbK\coloneqq \bbL\bbW $ is also of low rank. This is because of the temporal pattern in $\bbL$ in Fig. \ref{fig:P_observed}(a) leads to high correlation of the column differences of $\bbL$ as well. Clearly, the transformed matrix $\bbX$ satisfies the low-rank plus sparse structure, and its components $\bbK$ and $\bbD$ are related to the measurements in \eqref{sm_eqn}-\eqref{pmu_eqn} as
\begin{subequations}
\label{data_eqn_new} 
\begin{align}
\bbY&= (\bbK+\bbD)\bbU\bbA+\bbE_{y}, \label{sm_eqn_new} \\
\bbz^\T \bbW&=\mathbf{1}^{\mathsf{T}}(\bbK+\bbD)+\bbe_z^\T\bbW. \label{pmu_eqn_new} 
\end{align}
\end{subequations} 
%

Now the problem of recovering $\bbP$ becomes one of recovering a low-rank plus sparse matrix. One can introduce meaningful regularization terms to promote this structure. The nuclear matrix norm has been widely adopted for low-rank matrix recovery 
in problems such as matrix completion, subspace learning, and collaborative filtering; see e.g., \cite{recht2010guaranteed,candes2011robust,wright2009robust}. The matrix nuclear norm is defined as the sum of the singular values:
\begin{align}
\|\bbK\|_{*} \coloneqq \sum^{\min{\{N,T\}}}_{i=1} \sigma_i(\bbK) \label{nuc_norm},
\end{align}
where $\sigma_i(\cdot)$ denotes the $i$-{th} largest singular value. It is a convex function of the matrix input as shown in \cite[pg. 637]{Boyd}, because it is the dual function of the matrix spectral norm (or, the maximum singular value). 

To promote sparse $\bbD$, one can use the popular L1-norm regularization, as used in the fields of compressed sensing and sparse signal recovery; see e.g., \cite{Donoho_SIAM98,ct06tit,Donoho2006}. The L1-norm is defined as the sum of entry-wise absolute values, given by 
\begin{align}
\|\bbD\|_{1}\coloneqq\sum_{n,t}|D_{n,t}| \label{L1_D}. 
\end{align} 
Since the L1-norm is a tight convex relaxation of the L0 pseudo-norm (the number of nonzero entries), it has been shown to be able to efficiently find the sparse signal representation with performance guarantees. Note that the nuclear norm in \eqref{nuc_norm} can be thought of as the L1-norm of the matrix singular values. 
Therefore, minimizing the 
nuclear norm can lead to fewer numbers of nonzero singular values, and thus a low-rank matrix solution.  

Using the two norms, one can formulate the matrix $\bbX$ recovery problem as
\begin{subequations}
\label{opt_form}
\begin{align}
  \min_{\bbK, \bbD}&~~ \|\bbK\|_{*}+\lambda\|\bbD\|_{1} &\label{opt_obj}\\
    \mathrm{subject~to}&~~ -\xi_y\leq\bbY-(\bbK+\bbD)\bbU\bbA\leq\xi_y\label{opt_c1}\\
     &~~ -\xi_z \leq \bbz^\T \bbW-\mathbf{1}^{\mathsf{T}}(\bbK+\bbD)\leq\xi_z\label{opt_c2}
\end{align}  
\end{subequations}
where $\lambda>0$ is a fixed weight coefficient to balance $\bbK$ and $\bbD$ and parameters $\xi_y,\xi_z>0$ are pre-determined error bounds. We will discuss the choice of $\lambda$ in Sec. \ref{sec:perf}. As for the error bounds, they can be set according to the meter accuracy for each type of measurement or even to account for potential modeling inaccuracy due to e.g., bad data or feeder losses. If only measurement noise is considered, then the infinity norm based error constraints in \eqref{opt_c1}-\eqref{opt_c2} correspond to uniformly distributed noise. This assumption is valid for practical systems as meter accuracy is specified by the maximum error percentage. One can use different error criteria such as the Frobenius norm for Gaussian distributed noise.  Moreover, the error bounds can be different 
for every measurement entry, since they would scale with the actual data due to instrumentation as discussed in Sec. \ref{sec:simulation}. The recovery problem \eqref{opt_form} is a convex problem 
that can be solved by generic convex optimization solvers. The computational complexity may grow fast with the matrix dimension, thereby calling for accelerated solutions such as alternating minimization \cite{lin2010augmented} or adaptive updates using subspace learning approaches \cite{vaswani2018robust}, which will be explored in the future to develop fast algorithms for solving \eqref{opt_form}. 

One main issue of the proposed recovery formulation \eqref{opt_form} is that the L1-norm regularization may penalize the magnitude of nonzero entries of $\bbD$. Hence, the solution $\hhatbbD$ tends to be smaller in magnitude than the actual values and is thus biased. We have observed through numerical studies that the presence of frequent HVAC activities during the summer days could make 
this issue worse. This is because the periodic HVAC activities could exhibit a certain level of temporal pattern, and thus are partially captured by $\bbL$. As a result, the output $\hhatbbD$ from \eqref{opt_form} is more likely to suffer from smaller magnitudes.  



To tackle this issue, we will develop a post-processing scheme based on the recovered support of $\hhatbbD$ from \eqref{opt_form}. Albeit the magnitude bias, the nonzero entries of $\bbD$ can be accurately identified by solving \eqref{opt_form}. Hence, one can use the solution $\hhatbbD$ to obtain the subset of nonzero entries in $\mathcal{M}=\{(n,t)|~|\hhatD_{n,t}|>0\}$. In the numerical tests, we use a small positive threshold to reflect the numerical accuracy of the zero entries.  
%
Using the estimated support in $\ccalM$, one can recast the recovery problem \eqref{opt_form} by neglecting the penalty term on $\bbD$. Instead, an additional constraint can be introduced to directly set all other entries not in $\ccalM$ to be zero. With the given $\ccalM$, the post-processing problem is formulated as 
%
\begin{subequations}
\label{post_opt_form}
\begin{align}
    \min_{\bbK, \bbD}&~~ \|\bbK\|_{*} \label{post_opt_obj}\\
        \mathrm{subject~to}  &~~ D_{n,t} = 0, ~~\forall(n,t)\not\in\ccalM \label{post_opt_c3}\\
        &~~ -\xi_y\leq\bbY-(\bbK+\bbD)\bbU\bbA\leq\xi_y\label{post_opt_c1}\\
     &~~ -\xi_z \leq \bbz^\T \bbW-\mathbf{1}^{\mathsf{T}}(\bbK+\bbD)\leq\xi_z.  \label{post_opt_c2}
\end{align}  
\end{subequations}
This way, the nonzero entries of $\bbD$ can be better estimated. The full matrix recovery algorithm including this post-processing step is tabulated in Algorithm \ref{alg:post_algo}.

\begin{algorithm}[t]
\caption{Recovering matrices $\bbK$ and $\bbD$} \label{alg:post_algo}
\begin{algorithmic}[1]
	\State \textbf{Input:} Smart meter data $\bbY$ and D-PMU data $\bbz$.
	\State \textbf{Output:} Estimated $\hat{\bbK}$ and $\hat{\bbD}$
	\State \textbf{Step 1:} Solve the problem \eqref{opt_form} to obtain the biased estimation $\hat{\bbK}^b$ and $\hat{\bbD}^b$.
	\State \textbf{Step 2:} Find the set of nonzero entries $\ccalM:=\{(n,t)|~|\hhatD_{n,t}|>0\}$.
	\State \textbf{Step 3:} Solve the problem \eqref{post_opt_form} using $\ccalM$ to obtain the updated estimates $\hat{\bbK}$ and $\hat{\bbD}$.
\end{algorithmic}
\end{algorithm}

\subsection{Recovery Performance}
\label{sec:perf}
This subsection discusses the recovery guarantees that could be achieved by problem \eqref{opt_form}, as related to the theoretical results from the RPCA work; see e.g., {\cite{candes2011robust,Xu2012outlier,wright2009robust}}.  
%
As mentioned earlier, the general RPCA framework deals with the similar low rank plus sparse matrix form, but assumes the full (or partial) observability of the full matrix itself. Hence, the theoretical results therein is not directly applicable to our problem setting where the unknown matrix is observed with dimensionality reduction of both column- and row-space [cf. \eqref{data_eqn_new}], at the ratios of $T/T_s$ and $N/1$, respectively. Nonetheless, the RPCA results can still provide insightful intuitions regarding the recovery conditions and the parameter settings in our framework.

Certain conditions on $\bbK$ and $\bbD$ are required in order to achieve accurate RPCA results {\cite{candes2011robust,Chandrasekaran2011rank,vaswani2018robust}}. Loosely speaking, the low rank component $\bbK$ cannot be sparse and the sparse component $\bbD$ cannot be of low rank. For $\bbK$, 
the column space spanned by either its left or right singular vectors needs to have low to almost zero in \textit{coherency}
with the identity matrix, thereby ensuring the singular vectors are not sparse. This implies that the temporal pattern of load profiles cannot be sparse itself. Meanwhile, for the sparse component $\bbD$, the location of its non-zero entries should be random with no periodic or correlated patterns. The randomness condition would promote $\bbD$ to be, ideally, full rank, while a non-periodic sparse pattern would make sure no frequency-induced low-rank component exists in $\bbD$. As it will soon become clear in Section \ref{sec:simulation}, the non-periodic condition is very important in the analysis of summer-time data, where frequent HVAC activities could be problematic in recovering the low-rank and sparse components of the load matrix. 

The other insight provided by the RPCA work is the choice of weight $\lambda$, used to balance the low-rank and sparse components. 
As mentioned in {\cite{candes2011robust,wright2009robust}}, the $\lambda$ value should be chosen according to the matrix dimension, as $\ccalO(1/\sqrt{T})$ if the dimension $T\gg N$. 
This setting can effectively balance the scaling of the two norms with respect to $T$, at around $\sqrt{T}$ and $T$, respectively. It will be used by numerical tests in Section \ref{sec:simulation}. 

\section{Simulation Results}
\label{sec:simulation}

This section presents the numerical results of recovering $\bbP$ obtained by solving problem \eqref{opt_form} and using 
Algorithm \ref{alg:post_algo} on 
a winter and 
summer data set 
respectively. For the winter data set, it turns out the solution to \eqref{opt_form} is
sufficiently good as there is no periodic HVAC activity that could potentially lead to biased $\bbD$. However, for the summer data set with high level of HVAC activities, the post-processing step in Algorithm \ref{alg:post_algo} turns out to be useful 
as it improves the separation of 
the correlated PV profiles from the sparse appliance events. 

The ground-truth data $\bbP$ is obtained from the PecanStreet's Dataport \cite{pecan} for the 30 residential homes shown in Fig. \ref{fig:P_observed}(a) for one winter day and similarly for one summer day. The active power demand data is at minute-level resolution. 
Only half of the houses have PVs installed and all the houses are located in the same neighborhood (Mueller, Austin). Only 6 houses have EV charging events. Based on $\bbP$, we synthetically generate the measurements in \eqref{sm_eqn}-\eqref{pmu_eqn} by adding random noise. Additionally, we assume that the smart meters record at 15 minute intervals. 
Using the American National Standard Institute (ANSI) C12.20 Standard \cite{ANSI}, we assume the smart meters installed at each residential home are rated at $\pm$0.2\% error accuracy. Accordingly, the entries of $\bbE_y$ are independently drawn from a uniform distribution based on this accuracy.
As for the fast aggregated measurement $\bbz$, we follow from the
D-PMU data-sheet \cite{PSL}, where 
the $\pm$0.01\% total vector error of the phasor measurement leads to 
active power measurement error within $\pm$0.02\%. Hence, the entries of $\bbe_z$ are drawn from independent uniform distributions using the $\pm$0.02\% accuracy.  

To solve the convex problem \eqref{opt_form}, we need to determine the value of the tuning parameter $\lambda$ and the error bounds $\xi_y$ and $\xi_z$. For all of the following test cases, a value of 0.05 was used for $\lambda$. This value is chosen based on $\ccalO(1/\sqrt{T})$ (cf. Sec. \ref{sec:perf}) and it has produced consistently good recovery results. 
The error bounds $\xi_y$ and $\xi_z$ are determined based upon the added noise level to the synthetic data, where $\xi_y$ is set to be $0.2\%|\bbY|$ while  $\xi_z$ is $0.02\%|\bbz|^\T|\bbW|$. 

\begin{figure}[t]
	{
		\centering
		\includegraphics[width=\linewidth]{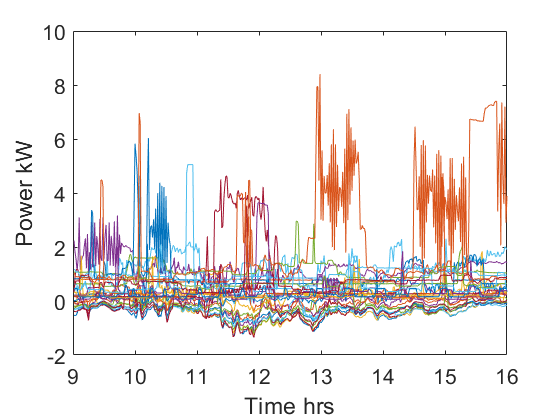}
		\centerline{(a)}
		\includegraphics[width=\linewidth]{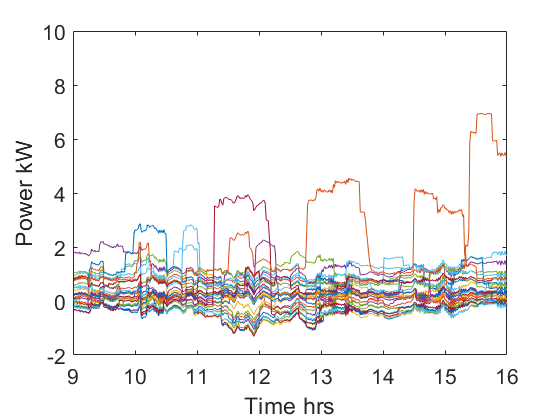}
		\centerline{(b)}
		\caption{Comparison between (a) the actual residential load profiles and (b) the recovered $\hat{\bbP}$ for the winter day-time loads in test case 1.}
		\label{fig:dec_day_Phat}}
\end{figure}

\begin{figure}[t!]
	{
		\centering
		\includegraphics[width=\linewidth]{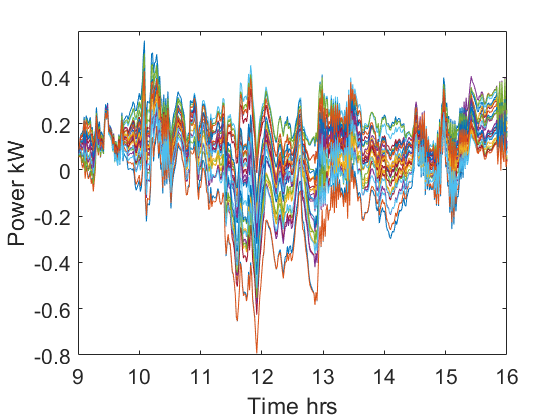}
		\centerline{(a)}
		\includegraphics[width=\linewidth]{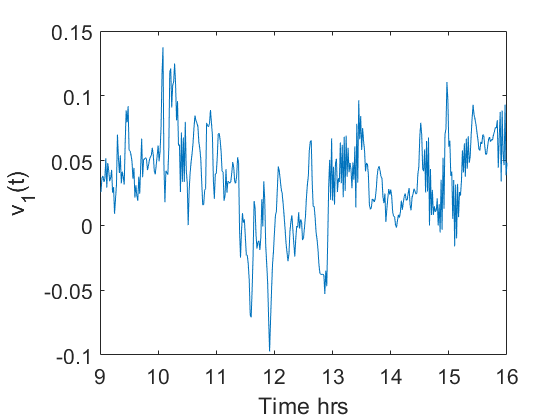}
		\centerline{(b)}
		\includegraphics[width=\linewidth]{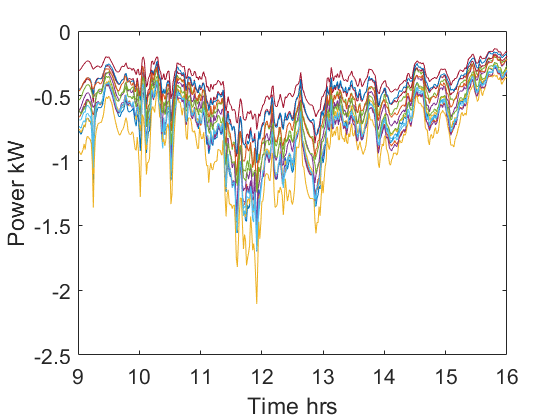}
		\centerline{(c)}
		\caption{Recovered winter solar irradiance pattern in test case 1: (a) the estimated $\hat{\bbL}$ for all 30 houses and (b) its first right singular vector, compared with with (c) the ground-truth PV output data for the 15 houses with PVs.}
		\label{fig:dec_irradiance}}
\end{figure}

To implement the convex problems \eqref{opt_form} and \eqref{post_opt_form}, we use the generic convex solver CVX \cite{cvx} in the MATLAB\textsuperscript{\textregistered} R2018a simulator, on
a laptop with Intel\textsuperscript{\textregistered} Core\texttrademark i7 CPU @ 2.10 GHz and 8 GB of RAM. Under this setting, the computational time for each problem is pretty reasonable, taking around 5-10 minutes to process a total of 7 hours of load data. We will develop accelerated solutions as mentioned in Sec. \ref{sec:recovery} in the future.  

\textit{1) Test Case 1 on winter day-time data:} We first test the winter data set for the day-time period from 9:00-16:00 to recover the solar irradiance pattern. Fig. \ref{fig:dec_day_Phat}(b) shows the recovered $\hat{\bbP}$ obtained by \eqref{opt_form}  which matches well with the ground-truth data in Fig. \ref{fig:dec_day_Phat}(a). 
Although $\hat{\bbP}$ is unable to capture the fast transients in the ground-truth data, it has included the major changes of dynamic load profiles such as the PV variations. 
To better illustrate the PV output recovery results, Fig. \ref{fig:dec_irradiance} plots the estimated $\hhatbbL$ and compares it with the ground truth data. Clearly, the estimated $\hat{\bbL}$ in Fig. \ref{fig:dec_irradiance}(a) is of very low rank, with its first right singular vector in Fig. \ref{fig:dec_irradiance}(b). As mentioned in Sec. \ref{sec:recovery} the temporal pattern is mainly due to the PV outputs. Although $\hhatbbL$ does not recover the actual ground-truth PV output, its first right singular vector has captured the main temporal pattern and can indicate the recovered PV output. It indeed matches well with the solar irradiance pattern present in the ground-truth PV outputs in Fig. \ref{fig:dec_irradiance}(c).
Since the winter data set does not contain HVAC activities, the low-rank component $\hhatbbL$ can be ideally separated from the sparse changes and it contains mainly the solar irradiance pattern. 

\begin{figure}[t!]
	{
		\centering
		\includegraphics[width=\linewidth]{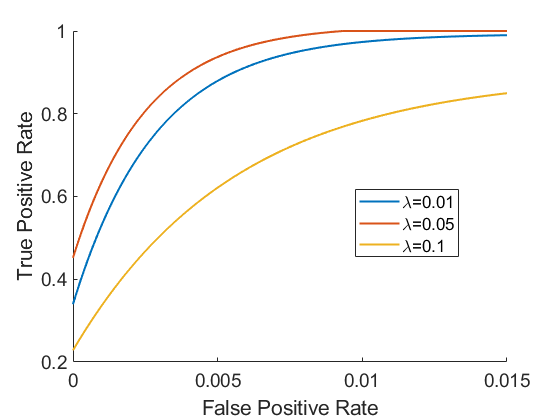}
		\caption{Receiving operating curve (fitted by an exponential function) for test case 2 showing the true positives versus false positives for detecting EV events.}
		\label{fig:roc}}
\end{figure}

\begin{figure}[t]
	{
		\centering
		\includegraphics[width=\linewidth]{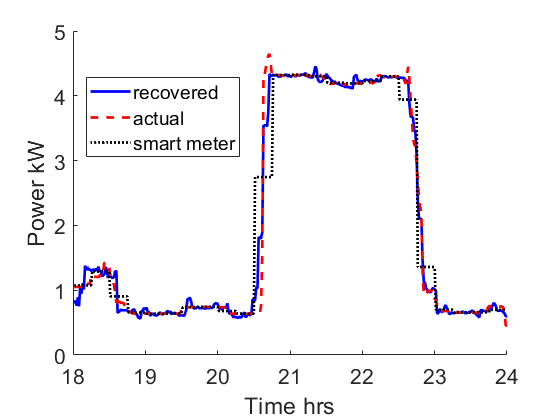}
		\caption{Recovered load data for one residential home with an EV charging event in test case 2.}
		\label{fig:dec_EV}}
\end{figure}

\begin{figure}[t!]
	{
		\centering
		\includegraphics[width=\linewidth]{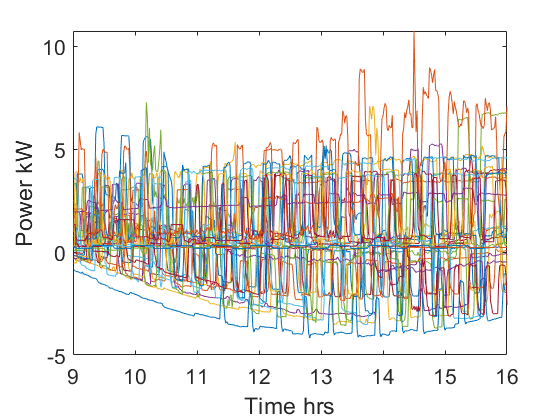}
		\caption{Ground-truth load profile for test case 3.}
		\label{fig:july_P}}
\end{figure}

\begin{figure}[t!]
	{
		\centering
		\includegraphics[width=\linewidth]{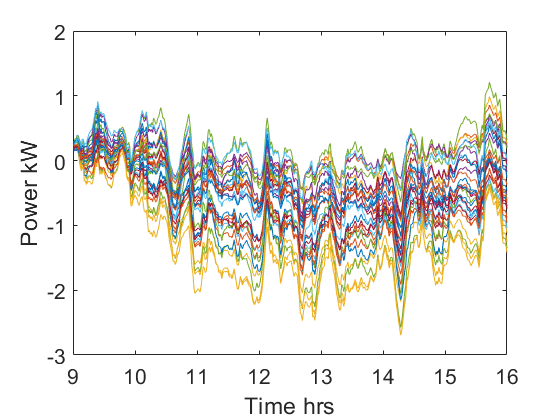}
		\centerline{(a)}
		\includegraphics[width=\linewidth]{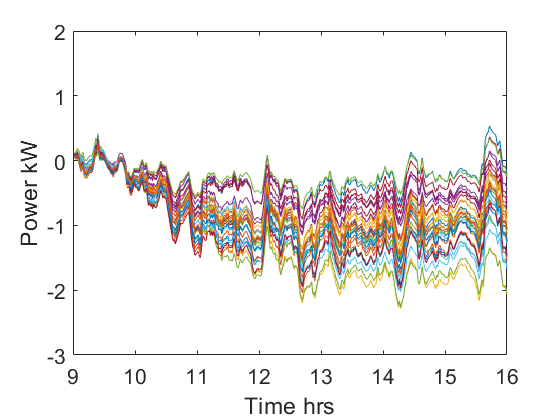}
		\centerline{(b)}
		\caption{Recovered $\hat{\bbL}$ from (a) the solution to \eqref{opt_form} and (b) Algorithm \ref{alg:post_algo} in test case 3.}
		\label{fig:july_Lhat}}
\end{figure}

\begin{figure}[t!]
	{
		\centering
		\includegraphics[width=\linewidth]{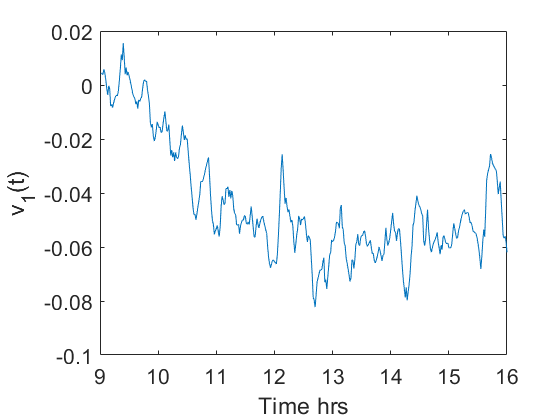}
		\centerline{(a)}
		\includegraphics[width=\linewidth]{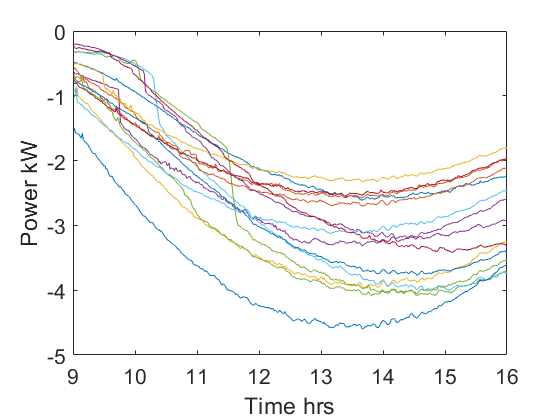}
		\centerline{(b)}
		\caption{Recovered solar irradiance pattern: (a) the first right singular vector of the estimated $\hat{\bbL}$ using Algorithm \ref{alg:post_algo}, and (b) the PV-only output data.}
		\label{fig:july_irradiance}}
\end{figure}

\begin{figure}[t!]
	{
		\centering
		\includegraphics[width=.9\linewidth]{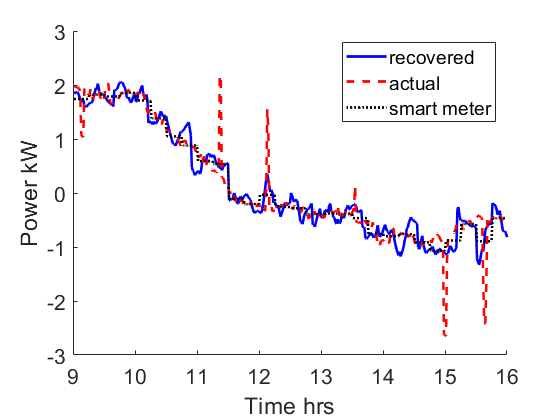}
		\centerline{(a)}
		\includegraphics[width=.9\linewidth]{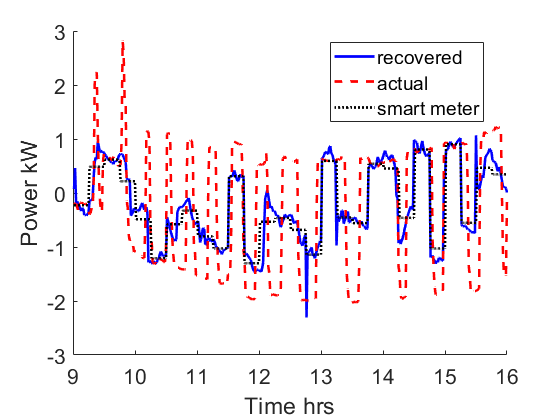}
		\centerline{(b)}
		\includegraphics[width=.9\linewidth]{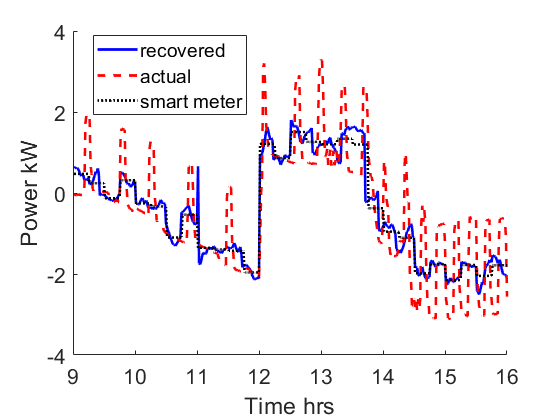}
		\centerline{(c)}
		\caption{Selected residential load profiles in test case 3 for a house with (a) no HVAC activity, (b) periodic HVAC activity, and (c) one EV charging event from 12:00-14:30.}
		\label{fig:july_recovery}}
\end{figure}

\textit{2) Test Case 2 on winter night-time data:} We further test the winter data set for the night time period from 18:00-24:00 to demonstrate the capability of recovered $\hat{\bbP}$ in identifying EV events. 
Fig. \ref{fig:roc} plots the receiving operating curve (ROC) of detected EV start/stop charging events as compared to the ground truth for three $\lambda$ values. Note that each ROC has been fitted with an exponential function. For each $\lambda$, the ROC is created by varying the detection threshold (as a percentage of EV power rating). As the EVs have larger power ratings than other appliances, the false positive (or false alarm) rate is very small. Fig. \ref{fig:roc} corroborates our recovery method's performance in EV identification, as all ROCs are close to perfect detection (top left corner of the plane). 
Additionally, the value $\lambda = 0.05$, as chosen for all our tests cases, is very competitive among the three.
One residential load is selected to demonstrate the recovery of an EV charging event 
occurring at around 20:30-23:00, as shown in Fig. \ref{fig:dec_EV}. Compared to the smart meter measurements which fail to indicate the exact EV charging start/finish time, our estimated profile can 
well match the actual minute-level profile.  

\textit{3) Test Case 3 on summer day-time data:} Lastly, we test the summer data set for the day-time period from 9:00-16:00 to demonstrate the impact of frequent HVAC activities.  
As shown in Fig. \ref{fig:P_observed}(b), there is a high level of HVAC activity in the summer data. The periodicity of the HVAC events would also manifest in a low-rankness of the load matrix, which would make it challenging to keep it solely in the sparse
component $\bbD$. To help improve the summer data recovery, we apply the post-processing step described in Algorithm \ref{alg:post_algo}.

Fig. \ref{fig:july_P} and Fig. \ref{fig:july_Lhat} plot respectively the 
ground-truth load profile and the recovered $\hat{\bbL}$ 
using both the solution from \eqref{opt_form} and Algorithm \ref{alg:post_algo}. Although the two estimates of $\hat{\bbL}$ 
look very similar,
the post-processing step has shown to improve the recovery of solar irradiance profile.  Without the post-processing step, there are some large, periodic transients present in the estimated $\hat{\bbL}$, as shown by Fig. \ref{fig:july_Lhat}(a). The transients are mainly due to the periodic HVAC activities and the penalty on the magnitude of $\hat{\bbD}$ in the objective of \eqref{opt_form}. After re-estimating 
$\hat{\bbL}$ and $\hat{\bbD}$ using \eqref{post_opt_form}, Algorithm \ref{alg:post_algo} is able to improve the recovery of 
the magnitude of the entries in the estimated support set $\ccalM$. Accordingly, Fig. \ref{fig:july_Lhat}(b) shows that the large transients have been reduced in $\hat{\bbL}$. To better show the recovery improvement 
in the PV profiles, Fig. \ref{fig:july_irradiance} compares the the first right singular vector of the post-processed $\hat{\bbL}$ to the ground-truth PV output data. The parabolic trend in the residential PV output data is well recovered by the low-rank component. 
Compared to the winter data results, the recovered $\hhatbbL$ here is affected by the periodic HVAC activities, and, therefore, exhibits some oscillation patterns of around 30-minute intervals as opposed to the smoother trend in the actual PV profiles.

Fig. \ref{fig:july_recovery} shows the recovered residential load profiles for three selected houses. The recovered load profiles captures the trend in the actual 
profiles. However, they are unable to match most of the fast HVAC events. Upon closer observation, we find out that the estimated profiles follow the smart meter data more often in the presence of periodic changes, as shown in Fig. \ref{fig:july_recovery}(b)-(c). Nonetheless, the proposed method can well capture the significant amount of change in the load profiles, such as EV charging as shown in Fig. \ref{fig:july_recovery}(c). Intuitively speaking, the performance degradation in recovering the summer time data is fundamentally due to the lack of observability in our system set-up. As there are increasing amount of transient events in the underlying load matrix, we need to use more measurements to be able to keep up with the unknown information. In other words, a single aggregated load profile 
provides insufficient amount of spatial diversity that 
is present in the unknown load data. Therefore, we should explore various types of data (voltage/current/reactive power) that D-PMUs can provide in addition to increasing the number of D-PMUs. This would require the incorporation of distribution feeder modeling and it is currently pursued to generalize the proposed load recovery framework. 


\section{Conclusions}
\label{sec:conclusion}

This paper presents a matrix recovery algorithm to enhance the spatio-temporal observability of residential loads by jointly utilizing both the smart meter and D-PMU data. Using the appropriate norm regularization, this problem is cast as a convex optimization one to promote the underlying low-rank and sparse change characteristics of the unknown load matrix. A post-processing procedure is developed as well to mitigate the estimation bias due to the regularization penalty. Numerical test results using real residential load data demonstrate that the recovery algorithm can effectively recover appliance activities and the PV output profiles. However, the presence of periodic HVAC loads would lead to some performance degradation in correctly identifying the sparse changes. 

We are currently exploring 
the use of various types of D-PMU data, in addition to increased number of D-PMUs,  for improved recovery performance. Furthermore, we plan to investigate accelerated and online solution methods that can be implemented efficiently and in real-time.

\section*{Acknowledgments}

This work has been supported by  the NSF Grant ECCS-1802319 and the DOE CEDS program.

\section*{References}
\vspace{-10pt}
\bibliographystyle{IEEEtran}
\bibliography{ref}
\end{document}